\documentclass[useAMS,usenatbib,onecolumn,usegraphicx]{mn2e}

\newcommand{\gsim}{\, \raisebox{-0.8ex}{$\stackrel{\textstyle >}{\sim}$ }}

\newcommand{\beq}{\begin{equation}}
\newcommand{\eeq}{\end{equation}}
\newcommand{\beqar}{\begin{eqnarray}}
\newcommand{\eeqar}{\end{eqnarray}}

\usepackage{ulem}
\usepackage{color}

\title[Constraining nuclear EOS \& NS structure]  
{Constraints on the nuclear equation of state and the neutron star 
structure from crustal torsional oscillations} 
\author[H. Sotani, K. Iida, \& K. Oyamatsu]
{Hajime Sotani$^1$ \thanks{E-mail:sotani@yukawa.kyoto-u.ac.jp},
Kei Iida$^2$, and
Kazuhiro Oyamatsu$^3$
\\
$^1$Division of Theoretical Astronomy, National Astronomical Observatory of Japan, 2-21-1 Osawa, Mitaka, Tokyo 181-8588, Japan\\
$^2$Department of Mathematics and Physics, Kochi University, 2-5-1 Akebono-cho, Kochi 780-8520, Japan\\
$^3$Department of Human Informatics, Aichi Shukutoku University, 2-9 Katahira, Nagakute, Aichi 480-1197, Japan}

\begin{document}
\maketitle
\label{firstpage}

\begin{abstract}
We systematically examine torsional shear oscillations of neutron star
crusts by newly taking into account the possible presence of the
phase of cylindrical nuclei. 
In this study, we neglect an effect of magnetic fields, under which 
the shear oscillations can be damped by the magnetic interaction.
First, by identifying the 
low frequency quasi-periodic oscillations (QPOs) observed in the 
soft-gamma repeaters (SGRs) as the fundamental torsional oscillations, 
we constrain the slope parameter of the nuclear symmetry energy, $L$, 
for reasonable values of the star's mass $M$ and radius $R$.  
Meanwhile, we find that the 1st 
overtone of torsional oscillations obtained for given $M$ and $R$ can 
be expressed well as a function of a new parameter 
$\varsigma\equiv (K_0^4 L^5)^{1/9}$, where $K_0$ is the incompressibility 
of symmetric nuclear matter.
Assuming that the lowest of the QPO frequencies above 500 Hz
observed in SGR 1806--20 comes from the 1st overtone, we can 
constrain the value of $\varsigma$. 
Then, for each neutron star model, such a value of $L$ as can be
obtained from the observed low frequency QPOs 
translates to the optimal value of $K_0$ via the above constraint 
on $\varsigma$.  Finally, its consistency with allowed values of $K_0$ 
from empirical giant monopole resonances leads to neutron star models 
with relatively low mass and large radius,  which are 
qualitatively similar to the prediction
in earlier investigations.  
This result suggests that $L\simeq 58$--73 MeV, 
even when uncertainties in the neutron superfluid density
inside the phase of cylindrical nuclei are allowed for.
\end{abstract}

\begin{keywords}
stars: neutron  -- equation of state -- stars: oscillations
\end{keywords}

\section{Introduction}
\label{sec:I}

Neutron stars, which are produced as stellar remnants of collapse 
of massive stars, exhibit states of matter in extreme conditions
where the density inside the star exceeds the normal nuclear density
significantly under the strong gravitational field, and the magnetic 
field inside/around the star can become extremely strong \citep{NS}. 
To extract the neutron star properties, asteroseismology is 
a powerful tool, just like seismology in the case of the Earth and 
helioseismology in the case of the Sun.  That is, by observing 
the spectra of electromagnetic waves and/or gravitational waves 
radiating from the objects, one would see the interior properties of the 
objects.  In fact, various eigenmodes could be excited in a neutron 
star in a manner that reflects the star's mass $M$ and radius $R$, 
the internal crystalline and superfluid properties, etc.\ \citep{VH1995}. 
For example, via the observations of gravitational waves, one would constrain 
$M$, $R$, and the equation of state (EOS) 
of neutron star matter 
\citep{AK1996,STM2001,SKH2004,SYMT2011,DGKK2013}.

Theoretically, the structure of 
a neutron star strongly depends on the uncertain EOS 
for neutron star matter, while the qualitative structure with the 
canonical mass of order $1.4M_\odot$ is fairly well established.  
In equilibrium, under the ocean composed of melted iron, the matter 
forms a lattice structure due to the Coulomb interaction and behaves as a 
solid.  This region is referred to as a crust.  As the density 
increases, the energy of matter in the lattice structure becomes larger than 
that of uniform matter.  Above such a critical density, the neutron star 
matter becomes uniform and behaves as a fluid.  This region is 
referred to as a core.  The critical density is predicted to 
lie roughly between the normal nuclear density and one third thereof 
in a manner that is dependent on the density dependence of the symmetry 
energy \citep{OI2007}.  
Furthermore, the existence of the non-spherical nuclear structure 
has been proposed at the basis of the crust \citep{LRP1993,O1993}, 
where the shape of non-spherical nuclei can change from spherical into 
cylindrical, slab-like, cylindrical-hole, and spherical-hole (bubble)
nuclei, as the density increases. These phases of non-spherical nuclei 
are often called pasta phases.  Anyway, the crust thickness is only
less than ten per cent of $R$, determined mainly by the stellar 
compactness $M/R$ and relatively weakly by such EOS parameters 
as the slope parameter of the nuclear symmetry energy $L$ and 
the incompressibility of symmetric nuclear matter $K_0$ \citep{SIO2017b}.

As far as asteroseismology in neutron stars is concerned, 
gravitational waves would convey helpful information, which is 
expected to be available in the near future.  Meanwhile, electromagnetic 
signals from oscillating neutron stars have been already 
detected as quasi-periodic oscillations (QPOs) in the afterglow of 
giant flares observed from soft-gamma repeaters (SGRs). Up to now, 
at least three giant flares were detected from SGR 0526--66, 
SGR 1900+14, and SGR 1806--20, together with the associated QPOs
\citep{B1983,I2005,SW2005,SW2006}.  In particular, several QPOs were
discovered at 28, 54, 84, and 155 Hz in SGR 1900+14 and at 18, 26, 
29, 92.5, 150, 626.5, and 1837 Hz in SGR 1806--20.  In addition to 
these QPOs observed in giant flares, a QPO of 57 Hz was also 
detected from SGR 1806--20 in less energetic bursts \citep{QPO2}. 
Recently, the possible finding of QPOs of 9.2 Hz in SGR 1806--20
and 7.7 Hz in SGR 1900+14 has been also suggested 
with Bayesian analysis \citep{PGSE2018}.
Since SGRs are generally identified as strongly magnetized neutron 
stars, the observed QPOs are likely to be associated with the
global oscillations. Such oscillations might be crustal torsional 
and/or magnetic oscillations, because the observed frequencies are
relatively lower than typical $p$-mode oscillations of order kHz.

Up to now, many studies of magnetic and elasto-magnetic
oscillations have been done by several groups
\citep{Sotani2007,Sotani2008a,Sotani2008b,Sotani2009,vHL2011,vHL2012,
CK2011,GCFMS2011,GCFMS2012,GCFMS2013a,GCFMS2013b,PL2013}. 
Given the magnetic field strength at the star's surface estimated from
the above SGRs \citep{K1998,H1999}, the magnetic fields should penetrate 
the stellar core \citep{Sotani2008b}.  Then, the magnetic oscillations 
can be coupled with the crustal torsional oscillations, which are 
responsible for so-called elasto-magnetic oscillations.  Details of this
coupling strongly depend on the strength of magnetic fields.  For 
example, the torsional oscillations would efficiently damp due to such 
coupling if the magnetic fields are stronger than $\sim 5\times 10^{13}$ 
G and dipolar \citep{GCFMS2011}.  Note, however, that the magnetic 
and elasto-magnetic oscillations are sensitive to the uncertain
magnetic field strength and distribution inside the star.  In addition, 
as long as the magnetic fields penetrate the stellar core, one 
requires information about the EOS of matter in the core, which is also 
uncertain, to construct the magnetic field structure
\footnote{There exits a recent attempt to constrain the EOS of matter in the core 
by comparing the elasto-magnetic oscillations to the observed QPO frequencies \citep{GCFMS2018}.}.
On the other hand, if 
one neglects the magnetic effects, one can focus on torsional 
shear oscillations that are confined in the crust.  Then,
one can avoid uncertainties in the magnetic field structure and in the 
core EOS, while constructing a canonical model that can reproduce the
observed QPO frequencies in a manner that is dependent on a limited number
of parameters characterizing the neutron star structure and the properties of
matter in the crust.  Here we thus neglect the magnetic effects and 
focus on purely crustal torsional oscillations.
In fact, assuming that the observed QPOs are identified as purely 
crustal torsional oscillations, one can constrain the EOS in the 
crust \citep{SA2007,SW2009,GNJL2011,PA2012,SNIO2012,SNIO2013a,SNIO2013b}. 
In addition, the QPOs could even give an imprint of the existence of 
pasta structures through the characteristic elastic properties
\citep{S2011,PP2016,SIO2017a}.

Most of the calculations of the eigenfrequencies of crustal torsional 
oscillations, however, have been done by ignoring the possible 
presence of nuclear pasta.  Since the thickness of the phases of 
non-spherical nuclei is at most about one per cent of $R$ 
\citep{SIO2017b}, such ignorance is seemingly reasonable.  In fact, the 
fundamental oscillations, of which the eigenfrequencies scale as the shear
speed divided by $2\pi R$, are insensitive to the possible presence of
nuclear pasta, although they are sensitive to $L$ via the $L$
dependence of the shear modulus \citep{SNIO2012}.  
On the other hand, the overtones, of which the eigenfrequencies scale as 
the shear speed divided by the crust thickness \citep{HC1980}, are
not necessarily so.  Furthermore, the slab-like nuclear structure behaves 
like a fluid rather than a crystal against long wavelength 
linear perturbations \citep{dGP1993,PP1998}.  Consequently, the crustal 
torsional oscillations can be separately excited inside the phases 
of spherical and cylindrical nuclei and inside the phases of 
cylindrical-hole and spherical-hole nuclei \citep{SIO2017a}. 
In this paper, we consider this possibility by newly examining the 
liquid-crystalline shear properties of cylindrical nuclei and then
systematically calculating the fundamental and 1st overtone 
eigenfrequencies of the crustal torsional oscillations.  By
comparing the results with the observed QPO frequencies, we will be
able to constrain not only $M$ and $R$, but also $L$, in a manner that
is almost independent of the poorly known neutron superfluid density
inside the phase of cylindrical nuclei.

In Sec.\ \ref{sec:II}, we construct the equilibrium configuration of a 
neutron star crust.  Section \ref{sec:III} is devoted to calculations of 
the shear modulus in the crust including the phase of cylindrical nuclei.  
In Sec.\ \ref{sec:IV}, we therefrom obtain the eigenfrequencies of torsional 
shear oscillations, which are then compared with the observed QPO frequencies.
Concluding remarks are given in Sec.\ \ref{sec:V}.
We use units in which $c=G=1$, 
where $c$ and $G$ denote the speed of light and the gravitational constant, 
respectively.

\section{Crust in equilibrium}
\label{sec:II}

It is generally accepted that the celestial objects, which are
responsible for SGRs and associated giant flares, are 
strongly magnetized neutron stars, i.e., magnetars.  Even so, since the 
magnetic energy is much smaller than the gravitational binding energy, one can 
neglect the contribution of the magnetic fields to the star's pressure
and density profiles.  The thermal energy can also neglected in describing
the star's structure.  Additionally, the observed rotational period 
from SGRs is generally very large. As a reasonable model for the
corresponding stars, therefore, we can safely consider a spherically 
symmetric neutron star, where the line element is given in spherical
coordinates by
\begin{equation}
  ds^2 = -e^{2\Phi(r)} dt^2 + e^{2\Lambda(r)} dr^2 + r^2 d \theta^2 + r^2\sin^2\theta d\phi^2.
\end{equation}
Here, the metric function, $\Lambda(r)$, is directly associated with 
the mass function, $m(r)$, via $e^{-2\Lambda}=1-2m/r$.

Now, neutron star models can be constructed by integrating 
the well-known Tolman-Oppenheimer-Volkoff (TOV) equations together with the 
zero-temperature EOS of matter in neutron 
stars.  Due to only a limited number of experimental and 
observational constraints on the EOS of dense nuclear matter, the 
EOS of neutron star matter has yet to be determined although 
numerous EOS models have been proposed up to now.  In particular, 
there are many uncertainties in the EOS of matter in the core, 
compared to that in the crust.  To avoid such uncertainties 
in the core EOS, therefore, we focus only on the crust region, where 
the equilibrium model is constructed by integrating the TOV equations from 
the the star's surface inward down to the bottom of the 
curst for a given set of $M$ and $R$ \citep{IS1997}.  In order to 
construct the crust in equilibrium, one has to prepare the EOS 
of matter in the crust, which is assumed to be composed of a mixture of
saturated nuclear matter (liquid) and pure neutron matter (gas) that is 
neutralized and beta equilibrated by a uniform gas of electrons.
In particular, we adopt the phenomenological EOS of nuclear matter
constructed in such a way as to reproduce empirical data for masses and 
charge radii of stable nuclei within the Thomas-Fermi approach \citep{OI2003}. 
By using the EOS of crustal matter obtained therefrom \citep{OI2007} 
(hereafter referred to as the OI-EOS), we systematically examine the 
dependence of the frequencies of torsional oscillations on the EOS parameters.

In the vicinity of the saturation density of symmetric nuclear matter, $n_0$, 
the bulk energy per baryon of uniform nuclear matter at zero temperature 
can be expressed as a function of baryon number density, $n_{\rm b}$, 
and neutron excess, $\alpha$, as
\begin{equation}
  w = w_0  + \frac{K_0}{18n_0^2}(n_{\rm b} - n_0)^2 + \left[S_0 + \frac{L}{3n_0}(n_{\rm b} - n_0)\right]\alpha^2, \label{eq:w}
\end{equation}
where $w_0$ and $K_0$ denote the saturation energy and incompressibility of 
symmetric nuclear matter, corresponding to $\alpha=0$ \citep{L1981}.  
Meanwhile, the coefficients affixed to the term of order 
$\alpha^2$, i.e., $S_0$ and $L$, are the parameters associated with the 
density-dependent symmetry energy $S(n_{\rm b})$. That is, $S_0$ is the symmetry
energy at $n_{\rm b}=n_0$, i.e., $S_0=S(n_0)$, while $L$ is the slope 
parameter given by $L\equiv 3n_0 (dS/dn_{\rm b})_{n_{\rm b}=n_0}$.   These five 
parameters, $n_0$, $w_0$, $K_0$, $S_0$, and $L$ are the parameters that
characterize the properties of nuclear matter around the saturation point. 
Through terrestrial nuclear experiments, whose data basically reflect
the properties of saturated nuclear matter that has a limited range of 
neutron excess, these parameters can be more or less constrained. 
Among them, $n_0$, $w_0$, and $S_0$ are relatively well-constrained, 
while the other two parameters, $K_0$ and $L$ are more difficult to determine. 
This is because one needs to obtain information for nuclear matter in a certain
range of density around the saturation point to determine the values of $K_0$ 
and $L$, which are higher order coefficients with respect to the change in 
density from $n_0$.  Within the extended Thomas-Fermi theory 
incorporating the bulk energy expression that reduces to Eq.\ (\ref{eq:w}) 
in the limit of $n_{\rm b}\to n_0$ and $\alpha\to0$, therefore, 
the OI-EOS was constructed by optimizing the values 
of $n_0$, $w_0$, and $S_0$ to reproduce experimental data for
masses and charge radii of stable nuclei for given values of $K_0$ and $L$
\citep{OI2007}.   The EOS parameter sets adopted in this study are 
shown in Table \ref{tab:EOS}, together with the transition density 
from spherical nuclei to cylindrical nuclei and that from cylindrical 
nuclei to slab-like nuclei (or to uniform matter).  Note that 
for $L\gsim 100$ MeV, no pasta was predicted to occur.

\begin{table*}
\centering
\caption{The EOS parameters adopted in this study and the corresponding 
transition density from spherical to cylindrical nuclei (SP--C) and that from 
cylindrical to slab-like nuclei (C--S).  The asterisk at the value of $K_0$ 
denotes the EOS model where cylindrical nuclei directly change to 
uniform matter.
}
\begin{tabular}{ccccccc}
\hline\hline
 & $K_0$ (MeV)  & $L$ (MeV) & $-y$ (MeV fm$^3$) & SP--C (fm$^{-3}$) & C--S (fm$^{-3}$) & \\
\hline
 & 180 & 5.7   & $1800$ & 0.06000 & 0.08665 &  \\
 & 180 & 31.0 &   $350$ & 0.05887 & 0.07629 &  \\
 & 180 & 52.2 &   $220$ & 0.06000 & 0.07186 &  \\
 & 230 & 7.6   & $1800$ & 0.05816 & 0.08355 &  \\
 & 230 & 42.6 &   $350$ & 0.06238 & 0.07671 &  \\
 & 230 & 73.4 &   $220$ & 0.06421 & 0.07099 &  \\
 & 280 & 54.9 &   $350$ & 0.06638 & 0.07743 &  \\
 & 280$^*$ & 97.5 &   $220$ & 0.06678 & 0.06887 &  \\
 & 360 & 12.8 & $1800$ & 0.05777 & 0.08217 &  \\
 & 360 & 76.4 &  $350$ & 0.07239 & 0.07797 &  \\
\hline\hline
\end{tabular}
\label{tab:EOS}
\end{table*}

     For calculating the frequencies of torsional oscillations in the
crust of a neutron star, one has to estimate the effective enthalpy density, 
$\tilde{H}$, that contributes to the oscillations.  In fact, such
frequencies are proportional to the shear speed defined as 
$v_s=({\mu/\tilde{H}})^{1/2}$ \citep{HC1980}, where $\mu$ denotes the shear 
modulus that will be described in the next section.  The effective 
enthalpy density can be obtained by subtracting the mass density of 
superfluid neutrons from the total enthalpy density, $H$, in equilibrium. 
Since the baryon chemical potential is given by $\mu_{\rm b}=H/n_{\rm b}$ 
at zero temperature, the effective enthalpy density can be
written as 
\begin{equation}
  \tilde{H} = \left(1-\frac{N_s}{A}\right)H,
\end{equation}
where $A$ is the baryon number in a Wigner-Seitz cell, while $N_s$ is the 
number of neutrons in a Wigner-Seitz cell that do not comove with protons 
in the nuclei \citep{SNIO2013a,SNIO2013b,SIO2017a}.

Since we will confine ourselves to spherical and cylindrical nuclei
in this work, we may assume that $N_s$ comes solely from a 
part of the dripped neutron gas.  Even under this assumption,  
it is uncertain how much fraction of dripped neutrons behave as a superfluid
\citep{CCH2005,Chamel2005,Chamel2012,WP2017}.  We thus introduce 
a new parameter $N_s/N_d$ with the number of dripped neutrons in the 
Wigner-Seitz cell, $N_d$.  Note that $N_s/N_d=0$ and 1 are the extreme 
cases, i.e., for $N_s/N_d=0$ all the dripped neutrons comove with the protons 
and $\tilde{H}$ is equivalent to $H$, while for $N_s/N_d=1$ all the dripped 
neutrons behave as a superfluid and do not participate in
the oscillations.  The value of $N_s/N_d$ in a realistic situation depends 
on the baryon density inside the star through the Bragg scattering 
of the dripped neutrons off the underlying crystalline structure.
In practice, $N_s/N_d$ in the phase 
of spherical nuclei becomes around 10--30 per cent for 
$n_{\rm b}\sim 0.01$--$0.4n_0$, according to band calculations that ignore
the effect of nonzero pairing gap \citep{Chamel2012}.  On the other hand, 
$N_s/N_d$ in the phase of cylindrical nuclei is more uncertain and more 
complicated, because the entrainment in this layer would be anisotropic 
with respect to the orientation of cylindrical nuclei. 
To obtain the isotropic effective mass of dripped neutrons, 
it is often assumed that the polycrystalline orientation of 
cylindrical nuclei is random.  Here again, according to the 
band calculations by \cite{Chamel2005,CCH2005} in the absence of the
effect of nonzero paring gap, the isotropic effective mass becomes 
$m_*/m=1.185$ at the $n_{\rm b}=0.06$ fm$^{-3}$, which corresponds to 
$N_s/N_d$ of $84.4$ per cent.  This may suggest that as compared 
to the phase of spherical nuclei, a much greater part of the dripped 
neutrons behave as a superfluid.  The effect of nonzero paring gap,
however, acts to reduce entrainment due to the band effect in a manner that 
is dependent on the ratio of the pairing and band gaps and on the dimension 
of the crystalline structure \citep{WP2017}.  In the present analysis,
for simplicity, we set $N_s/N_d$ in the phase of spherical nuclei 
to the results obtained by \cite{Chamel2012}, as in 
\cite{SNIO2013a,SNIO2013b}, while we consider $N_s/N_d$ in the phase of 
cylindrical nuclei as a free parameter that ranges $0\le N_s/N_d \le 1$. 
In particular, we will consider the extreme cases, i.e., $N_s/N_d=0$ and 1 
in the phase of cylindrical nuclei.  To describe crustal torsional
oscillations, therefore, we have to set the values of two stellar 
parameters, i.e., $M$ and $R$, two EOS parameters, i.e., $L$ and $K_0$, and 
$N_s/N_d$ in the phase of cylindrical nuclei.

\begin{figure}
\begin{center}
\includegraphics[scale=0.5]{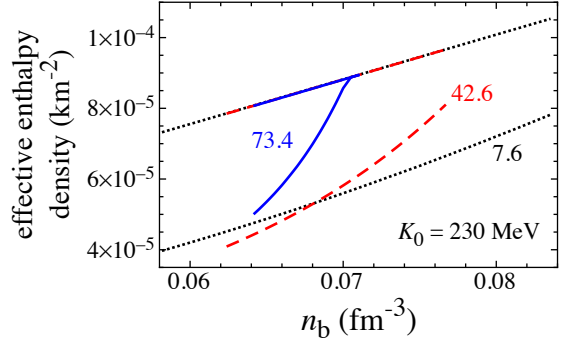} 
\end{center}
\caption{
Effective enthalpy density ($\tilde{H}$) in the phase of cylindrical nuclei.  
The dotted, dashed, and solid lines correspond to the results obtained
for $L=7.6$, $42.6$, and $73.4$ MeV, respectively, by fixing the value 
of $K_0$ to $230$ MeV.  For each EOS model, the upper and lower lines denote 
the effective enthalpy with $N_s/N_d = 0$ and 1. 
}
\label{fig:Hnb}
\end{figure}

\section{Shear modulus}
\label{sec:III}

The shear modulus is another important parameter 
for description of the crustal torsional oscillations. 
For a bcc lattice of spherical nuclei the effective shear modulus 
has been derived as
\begin{equation}
  \mu_{\rm sp} = 0.1194\frac{n_i(Ze)^2}{a},     \label{eq:musp}
\end{equation}
where $n_i$ is the number density of nuclei with the charge number of $Z$, 
and $a$ is the radius of the Wigner-Seitz cell, i.e., $1/n_i=4\pi a^3/3$ 
\citep{OI1990,SHOII1991}.  We remark that this shear modulus 
was derived by assuming that each nucleus is a point particle and by
taking average over all possible wave vectors of displacements in such a
way as to be relevant for polycrystalline matter with randomly oriented 
crystallites.  In the present analysis we simply adopt this 
traditional formula for the shear modulus in the phase of spherical nuclei, 
although modified versions of the shear modulus that 
allow for electron screening and more realistic but still randomly 
oriented polycrystalline 
nature \citep{KP2013,KP2015} are also proposed.  These 
modifications could act to reduce the effective shear modulus roughly 
by 30 per cent, while the effect of nonzero pairing gap could lead to
reduction of the effective enthalpy density by the same order or 
even larger than that \citep{WP2017}.  Fortunately, these two uncertain
factors would counteract with each other in evaluations of the torsional
oscillation frequencies, but eventually, full account of these factors 
would be desired.

Let us now move on to the shear modulus in the phase of cylindrical 
nuclei.  In this region, the equilibrium configuration of cylindrical
nuclei is likely to be a two-dimensional triangular lattice. 
According to \cite{dGP1993,PP1998}, who assumed that the direction of
cylindrical axes is locally $z$-direction and that the two-dimensional 
displacement of $(u^x,u^y)$ arises on the plane normal to the 
$z$-direction, the energy due to the deformation, $E_{\rm d}$, is given by 
\begin{eqnarray}
  E_{\rm d} &=& \frac{B_1}{2}\left(\frac{\partial u^x}{\partial x} + \frac{\partial u^y}{\partial y}\right)^2
     + \frac{C}{2}\left[\left(\frac{\partial u^x}{\partial x} - \frac{\partial u^y}{\partial y}\right)^2
                              + \left(\frac{\partial u^x}{\partial y} + \frac{\partial u^y}{\partial x}\right)^2\right]
     + \frac{K_3}{2}\left[\left(\frac{\partial^2 u^x}{\partial z^2}\right)^2 + \left(\frac{\partial^2 u^y}{\partial z^2}\right)^2\right] \nonumber \\
  &+& B_2 \left(\frac{\partial u^x}{\partial x} + \frac{\partial u^y}{\partial y}\right)
            \left[\left(\frac{\partial u^x}{\partial z}\right)^2 + \left(\frac{\partial u^y}{\partial z}\right)^2\right]
     + \frac{B_3}{2}\left[\left(\frac{\partial u^x}{\partial z}\right)^2 + \left(\frac{\partial u^y}{\partial z}\right)^2\right]^2,
\end{eqnarray}
where the terms with $B_1$, $C$, and $K_3$ are associated with the uniform 
transverse compression or dilation, transverse shear, and bending, 
respectively.  On the other hand, the terms with $B_2$ and $B_3$ are 
higher-order terms with respect to the displacement of $(u^x,u^y)$, 
which are negligible in the linear analysis as adopted here.  
For the torsional oscillations, therefore, only the coefficient $C$ 
is relevant and acts as a shear modulus.

     In order to estimate the value of $C$, one should calculate the change 
of energy by imposing an appropriate perturbation for transverse shear.  
In this case, $E_{\rm d}$ corresponds to the change in the Coulomb 
energy due to transverse shear that does not involve deformation of 
the cross section of each cylinder.  Let us now set the Coulomb energy per 
unit volume, $E_{\rm Coul}$, for the equilibrium phase of cylindrical nuclei.  
By calculating the spacially averaged Coulomb energy per unit volume, 
$\langle E_{\rm Coul}\rangle$, in the presence of the above perturbation
with respect to the displacement of $(u^x,u^y)$
and by identifying $E_{\rm d}$ with $\langle E_{\rm Coul}\rangle - E_{\rm Coul}$, 
one can estimate the value of $C$.  In practice, $C$ has been 
approximately estimated as
\begin{equation}
  C = E_{\rm Coul} \times 10^{2.1(w_2-0.3)},
\end{equation}
for a relevant range of the volume fraction, $w_2$, occupied by 
cylindrical nuclei \citep{PP1998}, 
where $w_2$ is defined as $w_2\equiv (R_p/R_c)^2$ 
with the radius of the cross section of a cylindrical nucleus $R_p$ and 
the corresponding Wigner-Seitz radius $R_c$.  In addition, given polycrystalline 
matter with randomly oriented crystallites as in the case of spherical 
nuclei\footnote{Orientations of crystallites might be affected
not only by gravitational and magnetic fields, but also by inhomogeneities in 
the local environment prior to nucleation.  In this study, however, we 
simply consider the case of randomly oriented crystallites.},
average of $C$ over all possible wave vectors of 
displacements leads to an effective shear modulus, $\mu_{\rm cy}$, as
\begin{equation}
  \mu_{\rm cy} = \frac{2}{3}C.   \label{eq:mucy}
\end{equation}
In the present analysis, we adopt this type of shear modulus in the 
phase of cylindrical nuclei for calculations of the eigenfrequencies of
torsional oscillations.  To estimate the effective shear modulus 
$\mu_{\rm cy}$ at given baryon density, we utilize the 
corresponding value of $E_{\rm Coul}$ that was obtained when the OI-EOS 
was constructed within the Thomas-Fermi model.  We remark that 
in the liquid drop model $E_{\rm Coul}$ is given by 
\begin{equation}
  E_{\rm Coul} = \frac{\pi}{2} (\rho_p R_p)^2 w_2\left[\ln\left(\frac{1}{w_2}\right)-1+w_2\right],
\end{equation}
where $\rho_p$ is the charge density in a cylindrical liquid drop, i.e., 
$\rho_p\equiv en_p$ with the local proton number density, $n_p$ \citep{RPW1983}.

Additionally, the phase of slab-like nuclei is predicted to occur just 
below the phase of cylindrical nuclei in a manner that is dependent 
on the EOS parameters.  The elastic properties in the phase of slab-like 
nuclei have been also discussed in \cite{dGP1993,PP1998}, where they showed 
that the energy-change due to the deformation is of higher order in 
the displacement.  That is, the phase of slab-like nuclei behaves as a fluid 
for long-wavelength torsional motion as can be examined in 
the linear analysis.  Even if additional pasta phases occur just below 
the phase of slab-like nuclei, therefore, the torsional oscillations 
are confined in the region of spherical and cylindrical nuclei, which can be 
analyzed independently of the torsional oscillations in the region of 
cylindrical-hole and spherical-hole nuclei \citep{SIO2017a}.

\begin{figure}
\begin{center}
\includegraphics[scale=0.5]{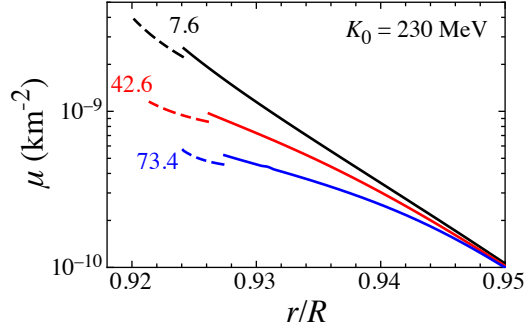} 
\end{center}
\caption{
Effective shear modulus in the phase of spherical nuclei ($\mu_{\rm sp}$) and 
in the phase of cylindrical nuclei ($\mu_{\rm cy}$), plotted by the solid 
and dashed lines, respectively, for stellar models with $1.4M_\odot$ and 
$12$ km.  Each line corresponds to the case of $L=7.6$, $42.6$, and 
$73.4$ MeV from top to bottom, where the value of $K_0$ is fixed to 
$230$ MeV.
}
\label{fig:shear}
\end{figure}

For this reason, we can focus on the oscillations that are confined
to the region of spherical and cylindrical nuclei, where the corresponding 
effective shear modulus is given by $\mu_{\rm sp}$, Eq.\ (\ref{eq:musp}), 
and $\mu_{\rm cy}$, Eq.\ (\ref{eq:mucy}), respectively.  In Fig.\ 
\ref{fig:shear}, we show the radial dependence of the effective shear 
modulus in the region of spherical and cylindrical nuclei for the stellar 
model with $1.4M_\odot$ and 12 km, which is obtained for three sets of the
EOS parameters.  We remark that the shear modulus 
would be graphically the same among the three cases if extended up to the 
star's surface.  From this figure, one can observe that the effective shear 
modulus decreases discontinuously at the transition point
from spherical nuclei to cylindrical nuclei.  This is due to the sudden 
change of the crystalline structure \citep{A2014}.

\section{Torsional oscillations and comparison with observed QPOs}
\label{sec:IV}

     In order to determine the frequencies of torsional oscillations in 
the crust of a spherical neutron star, we consider a linear analysis on 
the star's equilibrium configuration.  Since the torsional oscillations 
are of axial type and hence do not involve density variations,
we can safely adopt the relativistic Cowing approximation in which the 
metric perturbations are neglected.  Owing to the spherically symmetric 
background, the oscillations are described by one perturbation variable, 
i.e., the Lagrangian displacement (${\cal Y}$) of a given matter 
element in the $\phi$ direction.  The differential equation 
obeyed by this variable can be derived by linearizing the relativistic 
equation of motion as 
\begin{equation}
  {\cal Y}'' + \left[\left(\frac{4}{r} + \Phi' - \Lambda'\right)+\frac{\mu'}{\mu}\right]{\cal Y}' 
      + \left[\frac{\tilde{H}}{\mu}\omega^2e^{-2\Phi} - \frac{(\ell+2)(\ell-1)}{r^2}\right]e^{2\Lambda}{\cal Y}=0,
      \label{eq:perturbation}
\end{equation}
where $\tilde{H}$ is the effective enthalpy given in section \ref{sec:II}, 
the prime denotes the differentiation with respect to $r$, and $\omega$ 
denotes the angular frequency of torsional oscillations, of which 
the value will be determined from boundary conditions to be specified below 
\citep{ST1983}.  We notice that $\omega$ is associated with the  
frequency $f$ of torsional oscillations via $\omega=2\pi f$.  Since we 
consider the eigenmodes of torsional oscillations inside the phases of 
spherical and cylindrical nuclei in the present analysis, we impose 
relevant boundary conditions: At the star's surface, the torque 
vanishes, while at the base of the phase of cylindrical nuclei, 
the traction force vanishes.  Both conditions can be 
expressed as ${\cal Y}'=0$ \citep{ST1983,Sotani2007}.  Additionally, we 
impose a junction condition at the interface between the phase of 
spherical nuclei and the phase of cylindrical nuclei, which has to be 
a continuous traction condition, i.e.,
\begin{equation}
  \mu_{\rm sp}{\cal Y}' = \mu_{\rm cy}{\cal Y}'.
\end{equation}
Furthermore, since we can choose an arbitrary amplitude of torsional 
oscillations in Eq.\ (\ref{eq:perturbation}), we adopt the amplitude at the 
star's surface as a unit length.  Then, the problem to solve 
becomes an eigenvalue problem with respect to $\omega$.  Hereafter, we use 
the notation, ${}_nt_\ell$, in expressing the torsional 
eigenfrequencies with the angular index $\ell$ and the number $n$ 
of radial nodes in the eigenfunction.

\subsection{Fundamental oscillations}
\label{sec:IV-a}

\begin{figure}
\begin{center}
\begin{tabular}{cc}
\includegraphics[scale=0.5]{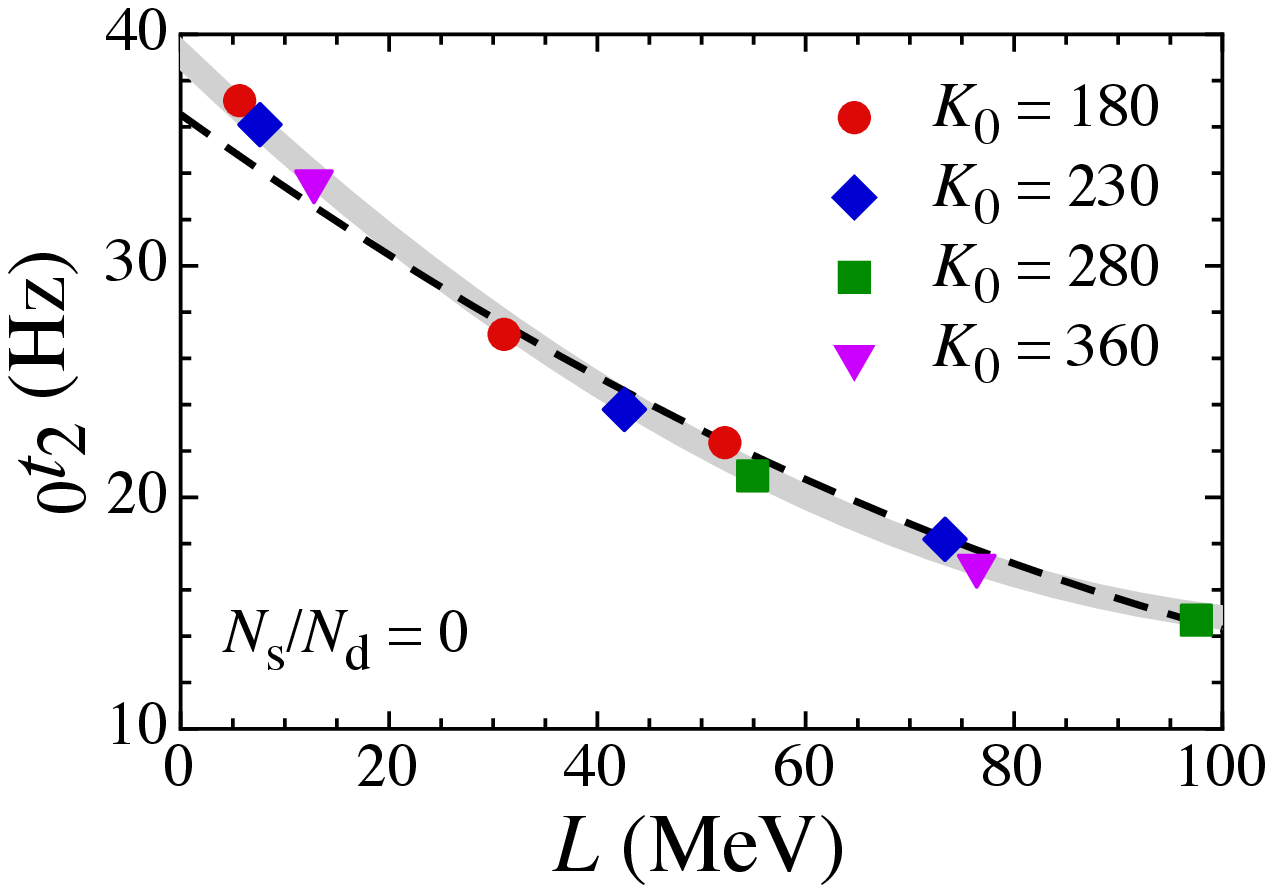} &
\includegraphics[scale=0.5]{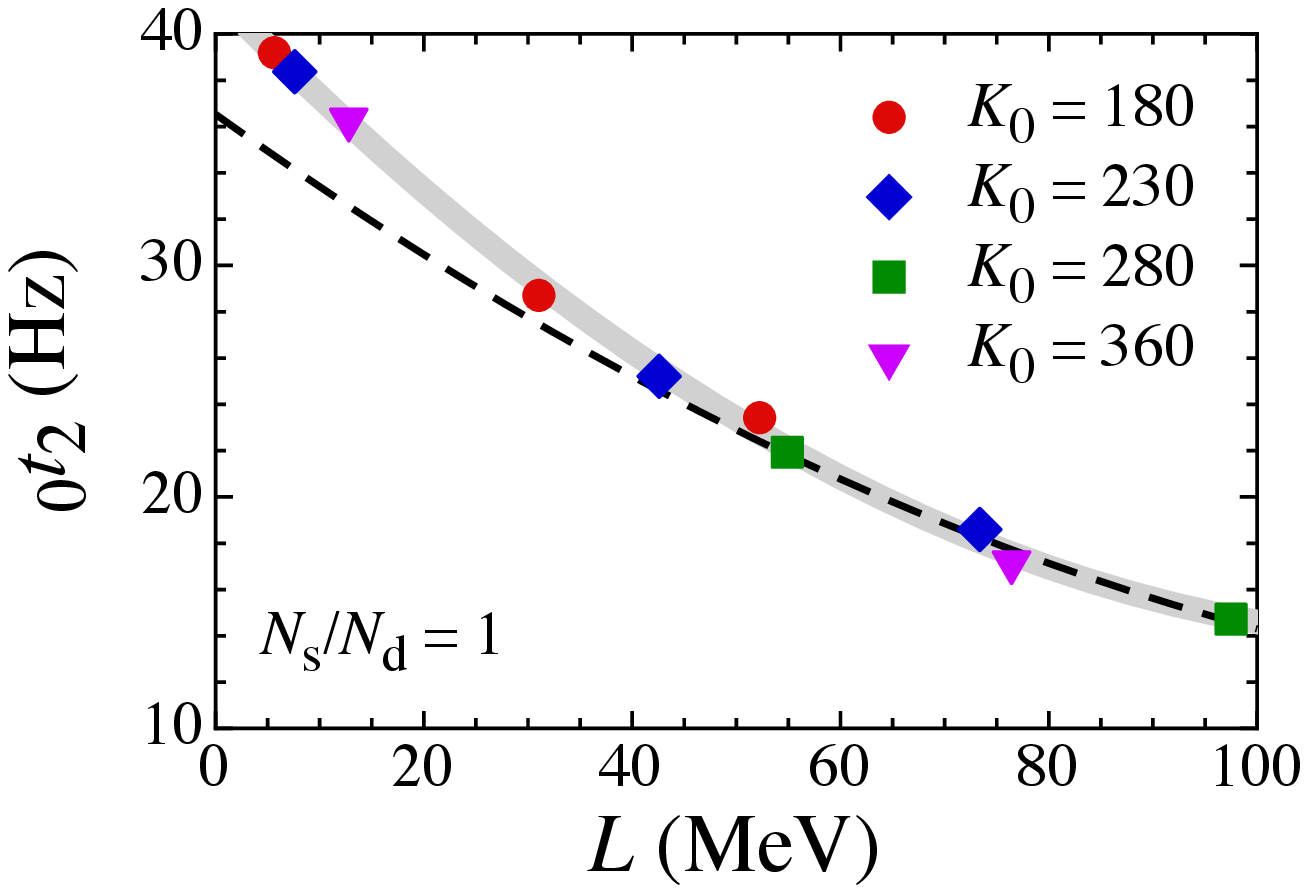} 
\end{tabular}
\end{center}
\caption{
The fundamental frequencies ${}_0t_2$ of the $\ell=2$ torsional oscillations
globally excited in the phases of spherical and cylindrical nuclei, 
which are calculated for various EOS models and for stellar models with 
$M=1.4M_\odot$ and $R=12$ km. The results are plotted in symbols 
as a function of $L$ in the case of $N_s/N_d=0$ (left) and $N_s/N_d=1$ 
(right) in the phase of cylindrical nuclei.  In both panels, the thick 
solid line denotes the fitting formula given by Eq.\ (\ref{eq:0t2L}), while 
the dashed line denotes the same kind of formula for the oscillations 
not involving the phase of cylindrical nuclei.
}
\label{fig:0t2-M14R12}
\end{figure}

First, we examine the fundamental frequencies of torsional oscillations, i.e., 
${}_0t_\ell$, in a manner that is dependent on the EOS parameters $L$ and 
$K_0$.  A similar analysis for the fundamental crustal torsional 
oscillations confined in the phase of spherical nuclei has been already 
done \citep{SNIO2012,SNIO2013a,SNIO2013b,S2014,SIO2016,S2016}, where it 
was shown that the fundamental frequencies of torsional oscillations are 
almost independent of the value of $K_0$.  In the presence of cylindrical
nuclei, however, the earlier calculations are not always realistic because the 
cylindrical phase has nonzero shear modulus as shown in Fig.\ \ref{fig:shear}.
Now, we thus calculate the $\ell=2$ fundamental frequencies  
of torsional oscillations that are globally excited in the phases of spherical 
and cylindrical nuclei.   The resultant frequencies, which are obtained 
for stellar models with $M=1.4M_\odot$ and $R=12$ km, various EOS 
models shown in Table \ref{tab:EOS}, and $N_s/N_d=0,1$ in the phase of 
cylindrical nuclei, are exhibited in Fig.\ \ref{fig:0t2-M14R12}.  From 
this figure, we can confirm that the $\ell=2$ fundamental frequencies 
of oscillations involving the phase of cylindrical nuclei are 
still almost independent of the value of $K_0$.  In practice, we find 
that the dependence of the $\ell=2$ fundamental frequencies on $L$ can be 
expressed as
\begin{equation}
  {}_0t_2 = c_2^{(0)} + c_2^{(1)} L + c_2^{(2)}L^2,  \label{eq:0t2L}
\end{equation}
where $c_2^{(0)}$, $c_2^{(1)}$, and $c_2^{(2)}$ are coefficients that are
determined by a fit to the calculations of ${}_0t_2$ for various $L$.
The values from this fitting formula are also plotted in Fig.\ 
\ref{fig:0t2-M14R12}, together with those from the same kind of fitting 
formula for the oscillations confined in the phase of spherical nuclei. 
By comparing these two cases, we find that the effect of the phase of 
cylindrical nuclei is generally small and acts to increase the 
fundamental frequencies of torsional oscillations only for small values 
of $L$ for which the phase of cylindrical nuclei is predicted to have a 
relatively large density range as shown in Table \ref{tab:EOS}.

In addition, 
we similarly calculate the $\ell$-th order fundamental frequencies of 
torsional oscillations, ${}_0t_\ell$, from various EOS models and find
that the dependence of ${}_0t_\ell$ on $K_0$ is negligible
for neutron star models ranging $M=1.4$--$1.8M_\odot$ and $R=10$--14 km. 
Thus, one can generally express ${}_0t_\ell$ as a function of $L$ as
\begin{equation}
  {}_0t_\ell = c_\ell^{(0)} + c_\ell^{(1)} L + c_\ell^{(2)}L^2,  \label{eq:0tlL}
\end{equation}
where $c_\ell^{(0)}$, $c_\ell^{(1)}$, and $c_\ell^{(2)}$ are coefficients that
can be determined from fitting to the calculations of ${}_0t_\ell$ 
as a function of the star's mass and radius.

\begin{figure}
\begin{center}
\begin{tabular}{cc}
\includegraphics[scale=0.5]{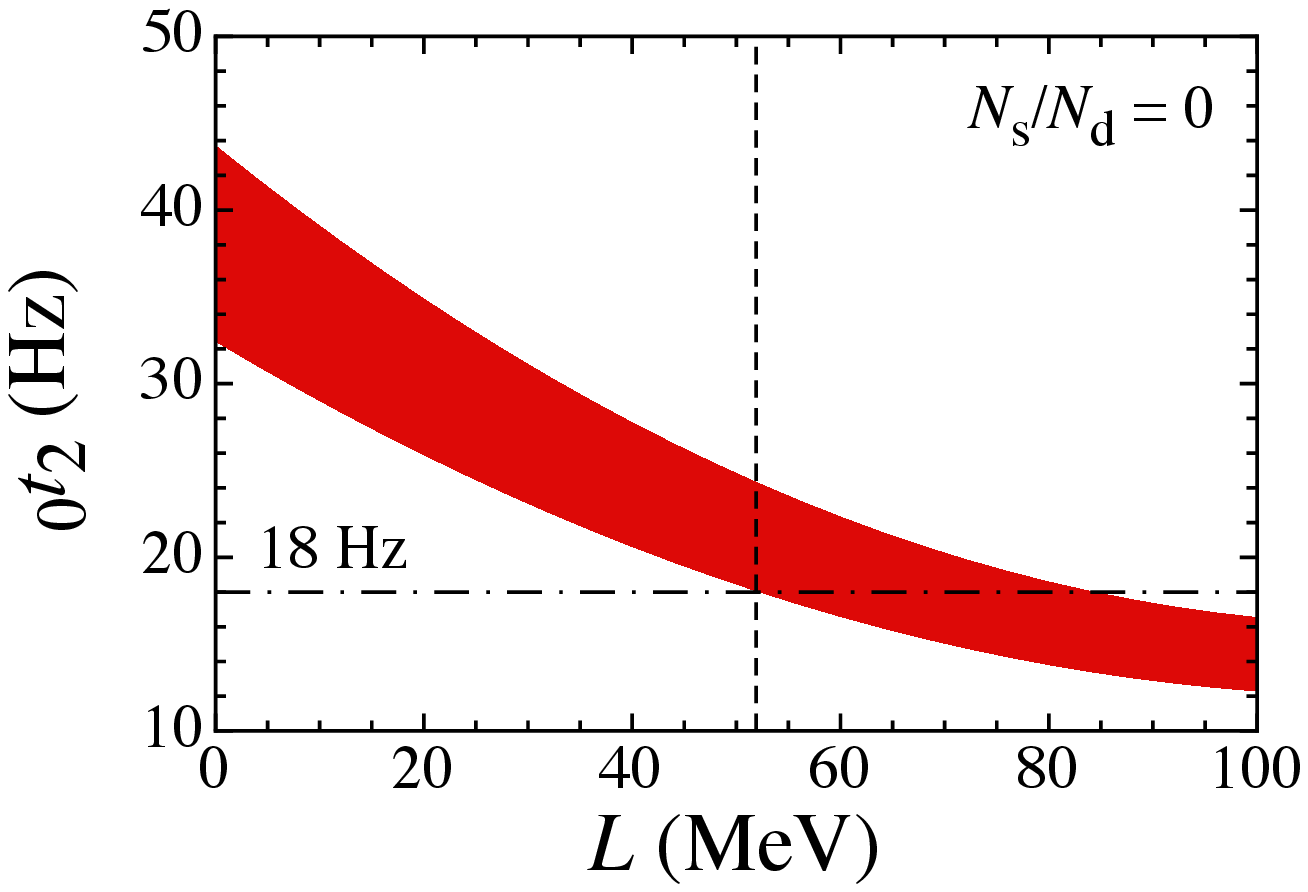} &
\includegraphics[scale=0.5]{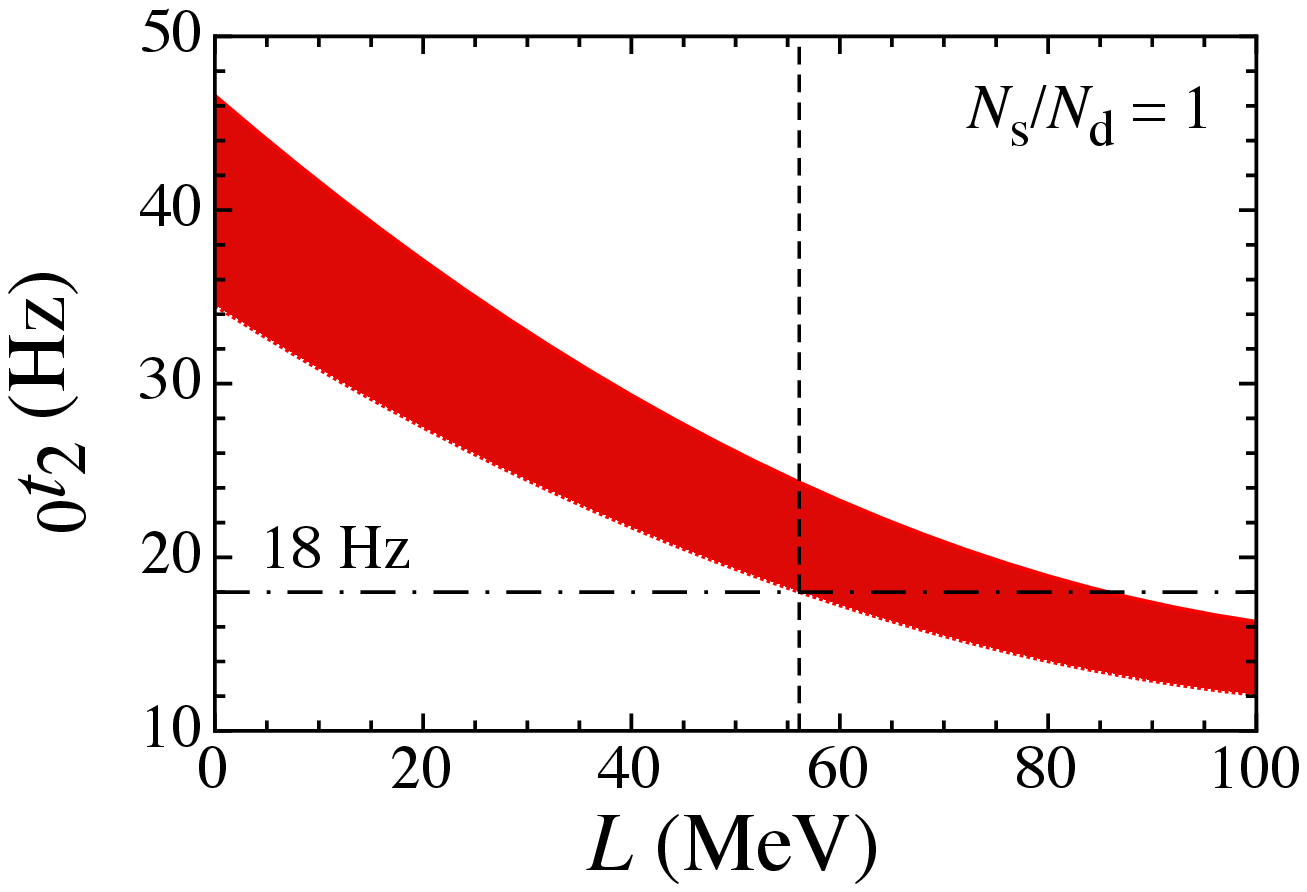} 
\end{tabular}
\end{center}
\caption{
The $L$ dependence of the eigenfrequency ${}_0t_2$ calculated for 
stellar models with $10 \le R\le 14$ km and $1.4\le M/M_\odot \le 1.8$ 
in the case of $N_s/N_d=0$ (left panel) and 1 (right panel) 
in the phase of cylindrical nuclei.  The horizontal dot-dashed 
line denotes the lowest frequency observed in the giant flare 
observed from SGR 1806--20, while the vertical dashed line 
corresponds to $L=51.9$ MeV (left panel) 
and $56.1$ MeV (right panel). 
}
\label{fig:L-0t2}
\end{figure}

Let us proceed to constrain the value of $L$ by comparing the $\ell=2$ 
fundamental frequencies of crustal torsional oscillations with the 
lowest QPO frequency observed in giant flares.  In Fig.\ \ref{fig:L-0t2}, the 
prediction of ${}_0t_2$ is shown for stellar models with 
$M=1.4$--$1.8M_\odot$ and $R=10$--14 km in the case of $N_s/N_d=0$ and 1 in 
the phase of cylindrical phase.  Note that ${}_0t_2$ is the lowest frequency 
among the torsional oscillations.  Given our interpretation of the QPO
frequencies in terms of crustal torsional oscillations, the lowest QPO 
frequency, i.e., 18 Hz, in SGR 1806--20 has to be higher than ${}_0t_2$.
We thus obtain a constraint of $L$ as $L \gsim 51.9$ MeV for $N_s/N_d=0$ 
and $L\gsim 56.1$ MeV for $N_s/N_d=1$.

One may further constrain the value of $L$ by identifying the observed 
QPOs as manifestations of various crustal torsional oscillations.  
In fact, we have shown the possibility of constraining $L$ by fitting 
the calculated fundamental torsional oscillations with various $\ell$ 
to the observed QPO frequencies lower than 100 Hz in our earlier 
investigations, where we neglected the effect of the phase of cylindrical 
nuclei \citep{SNIO2013a,SNIO2013b,SIO2016}.  In such an attempt, we found 
two possibilities of explaining the observed low frequency QPOs 
in terms of the crustal torsional oscillations, i.e., (i) 18, 26, 
29, 57, and 92.5 Hz in SGR 1806--20 correspond to the $\ell=3$, 4, 5, 
9, and 15 fundamental torsional oscillations, and 28, 54, and 84 Hz in 
SGR 1900+14 correspond to the $\ell=4$, 8, and 13 fundamental torsional 
oscillations. (ii) 18, 29, 57, and 92.5 Hz in SGR 1806--20 
correspond to the $\ell=2$, 3, 6, and 10 fundamental torsional 
oscillations, and 28, 54, and 84 Hz in SGR 1900+14 correspond to the
$\ell=3$, 6, and 9 fundamental torsional oscillations 
\footnote{In our previous studies where the effect of the phase of 
cylindrical nuclei was neglected, we adopted 30 Hz rather 
than 29 Hz as one of the observed QPO frequencies in SGR 1806--20. 
In fact, \cite{I2005} first reported the presence of the 30 Hz QPO.
After that, however, \cite{SW2006} concluded by reanalyzing the 
archival {\it RXTE} data that the QPO frequency is closer to 29 Hz than 
30 Hz.  In the present analysis, therefore, we adopt 29 Hz as 
the QPO frequency.  As shown in text, 29 Hz is more preferable 
than 30 Hz in the second possibility of identifying the QPOs.}.
As for the first possibility, one can explain all the observed low frequency
QPOs in terms of the crustal torsional oscillations,
where the optimal value of $L$ becomes larger than 
100 MeV \citep{SNIO2013a},
but this constraint seems too large to be consistent with existing
nuclear experiments, which implies typically $L=30$--80 MeV 
\citep{Tsang2012,L2014,Newton2014,BB2016}.  Meanwhile, 
the constraint on $L$ resulting from the second possibility
is more or less consistent with the experiments, although one has to 
invoke an additional oscillation mechanism for explaining the 
unidentified 26 Hz QPO.  As a possible mechanism for that, 
we proposed torsional oscillations excited in the phase of bubble nuclei
\citep{SIO2017a}; strictly speaking, these torsional oscillations,
if occurring, would extend to the phase of cylindrical-hole nuclei, 
but would not affect the torsional oscillations in the phases of 
spherical and cylindrical nuclei because of vanishing shear 
modulus in the intervening phase of slab-like nuclei
Hereafter, we shall 
adopt the second possibility.  In addition to the consistency with the 
empirical constraint on $L$ as mentioned above, there are several reasons
for that.  One is that the two QPO frequencies, 26 and 29 Hz, are too 
close to make the first possibility satisfactory.  Another is that 
the $\ell=2$ torsional oscillation frequency confined in the phases of 
bubble and cylindrical-hole nuclei can reproduce the 26 Hz QPO given
uncertainty in the entrainment effect.

\begin{figure}
\begin{center}
\includegraphics[scale=0.5]{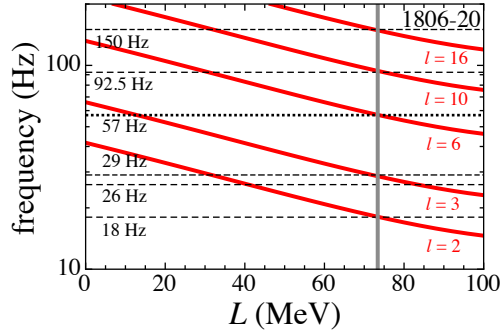} 
\end{center}
\caption{
Comparison of the low-lying QPO frequencies observed in 
SGR 1806--20 (horizontal dashed and dotted lines) 
and the fundamental frequencies 
of crustal torsional oscillations (solid lines) obtained as a function 
of $L$ for a neutron star model with $M=1.4M_\odot$, $R=12$ km, and 
$N_s/N_d=1$ in the phase of cylindrical nuclei.  The vertical 
thick solid line denotes the optimal value of $L$ that 
is consistent with the low frequency observed QPOs except 
the 26 Hz one. 
}
\label{fig:1806-M14R12Ns10}
\end{figure}

\begin{figure}
\begin{center}
\includegraphics[scale=0.5]{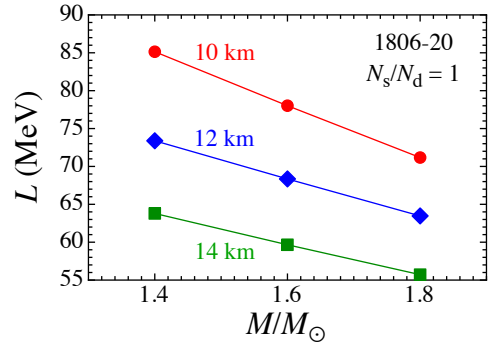} 
\end{center}
\caption{
The optimal values of $L$ that reproduce the low-lying 
QPO frequencies observed in SGR 1806--20 except 26 Hz in terms 
of the crustal torsional oscillations are plotted for various neutron 
star models with $N_s/N_d=1$ in the phase of cylindrical nuclei. 
}
\label{fig:1806-Ns10}
\end{figure}

As shown in Fig.\ \ref{fig:1806-M14R12Ns10}, the low frequency QPOs 
observed in SGR 1806--20 can in fact be explained in terms of
the fundamental frequencies of crustal torsional oscillations 
with the specific values of $\ell$ even by taking into account the 
presence of the phase of cylindrical nuclei. 
In particular, we find that the 150 Hz QPO can be explained 
in terms of the fundamental crustal torsional oscillations, 
simultaneously with the observed QPOs of frequencies lower than 
100 Hz except the 26 Hz one; the corresponding $\ell$ is 16.  
From this figure, 
the optimal value of $L$ is found to be $L=73.4$ MeV for
a typical neutron star model with $M=1.4M_\odot$ and $R=12$ km
and with $N_s/N_d=1$ in the phase of cylindrical nuclei.
Since the fundamental frequencies of the crustal torsional oscillations 
are predicted to decrease with $M$ and $R$, the optimal 
value of $L$ decreases with $M$ and $R$ as shown in Fig.\ 
\ref{fig:1806-Ns10}, where the range of $M=1.4$--$1.8M_\odot$ and 
$R=10$--14 km and the value of $N_s/N_d=1$ in the phase of cylindrical nuclei 
are assumed.  In this case, the optimal value of $L$ turns out to be in the 
range of $L=55.7$--85.1 MeV.

\begin{figure}
\begin{center}
\includegraphics[scale=0.5]{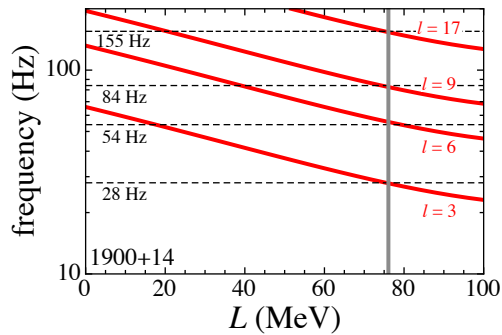} 
\end{center}
\caption{
Same as Fig.\ \ref{fig:1806-M14R12Ns10}, but for the QPO 
frequencies observed in SGR 1900+14.
}
\label{fig:1900-M14R12Ns10}
\end{figure}

\begin{figure}
\begin{center}
\includegraphics[scale=0.5]{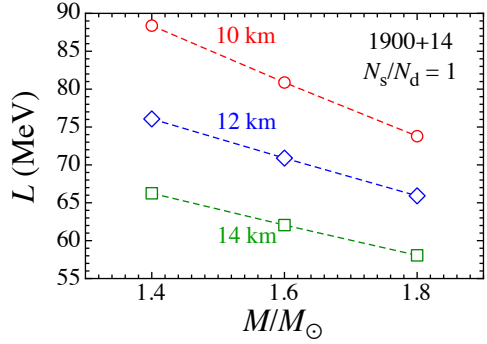} 
\end{center}
\caption{
Same as Fig.\ \ref{fig:1806-Ns10}, but for the low-lying QPO 
frequencies observed in SGR 1900+14.
}
\label{fig:1900-Ns10}
\end{figure}

In the same way, we find that the QPOs observed in 
SGR 1900+14 can be also identified as the fundamental 
crustal torsional oscillations, as shown in Fig.\ 
\ref{fig:1900-M14R12Ns10} for a typical neutron star model 
with $M=1.4M_\odot$, $R=12$ km, and $N_s/N_d=1$ in the phase 
of cylindrical nuclei. 
In particular, we find that the 155 Hz QPO can be identified 
as the $\ell=17$ fundamental crustal torsional oscillations, with
the same accuracy as the identifications of the other QPO frequencies.
In this case, the optimal value of $L$ 
is $L=76.1$ MeV.  Here again, the optimal value of $L$ 
changes with $M$ and $R$ as shown in Fig.\ \ref{fig:1900-Ns10}, 
where the range of $M=1.4$--$1.8M_\odot$ and $R=10$--14 km and the value 
of $N_s/N_d=1$ in the phase of cylindrical nuclei are assumed.
From this figure, we find that the optimal value of $L$ is in the 
range of $L=58.1$--88.4 MeV.

\begin{figure}
\begin{center}
\includegraphics[scale=0.5]{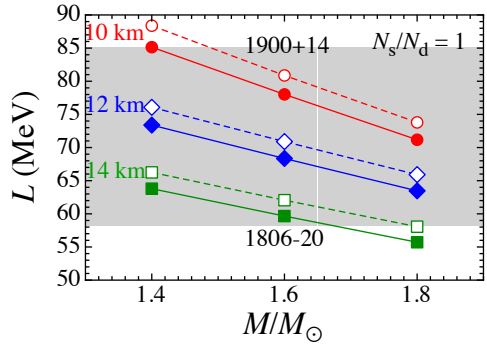} 
\end{center}
\caption{
The values of $L$ that simultaneously explain both 
of the low frequency QPOs observed in SGR 1806--20 (except 
the 26 Hz one) and SGR 1900+14 in terms of the 
fundamental torsional oscillations in the crust of a neutron star 
whose mass and radius range $1.4$--$1.8M_\odot$ and 10--14 km.
The painted region denotes such values of $L$ in the case of 
$N_s/N_d=1$ in the phase of cylindrical nuclei, while the symbols 
denote the optimal values of $L$ plotted in Figs.\ 
\ref{fig:1806-Ns10} and \ref{fig:1900-Ns10}.
}
\label{fig:fit-Ns10}
\end{figure}

\begin{figure}
\begin{center}
\includegraphics[scale=0.5]{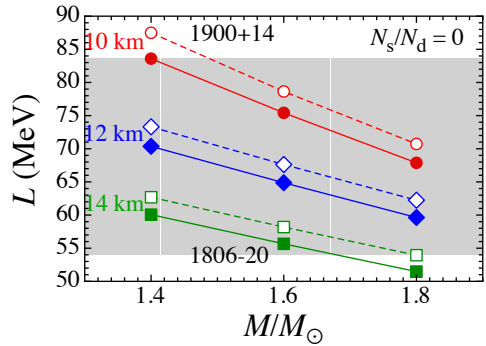} 
\end{center}
\caption{
Same as Fig.\ \ref{fig:fit-Ns10}, but with $N_s/N_d=0$ in the
phase of cylindrical nuclei.
}
\label{fig:fit-Ns00}
\end{figure}

Since the value of $L$ has to be uniquely determined, 
we can constrain the allowed values of $L$ in such a way as to 
simultaneously explain the low frequency QPOs observed in SGR 
1806--20 and in SGR 1900+14 in terms of the fundamental 
torsional oscillations in the crust of a neutron star of which
the mass and radius is in the range of $M=1.4$--$1.8M_\odot$ and 
$R=10$--14 km.  In the case of $N_s/N_d=1$ and $0$ in the
phase of cylindrical nuclei, as can be seen from Figs.\ 
\ref{fig:fit-Ns10} and \ref{fig:fit-Ns00}, such constraint on 
$L$ reads $L=58.1$--85.1 MeV and $L=53.9$--83.6 MeV, respectively.
This suggests that uncertainty in $N_s/N_d$ in the phase of 
cylindrical nuclei makes only a little difference in the 
constraint.  In fact, even with such uncertainty taken into
account, the allowed $L$ lies in the range of 
$L=53.9$--85.1 MeV.

\subsection{The 1st overtones}
\label{sec:IV-b}

Next, we turn to the properties of the 1st overtones of crustal 
torsional oscillations. Theoretically, the corresponding frequencies, 
${}_1t_\ell$, are considered to be associated with the crust thickness, 
$\Delta R$, as ${}_1t_\ell \propto v_s/\Delta R$ \citep{HC1980}, 
whereas $\Delta R$ in turn depends on the EOS parameters $L$ and 
$K_0$ mainly through the neutron chemical potential at the crust-core boundary 
\citep{SIO2017b}.  Via identification of the relatively high frequency 
QPOs observed in SGR 1806--20 as the overtones of crustal torsional 
oscillations, therefore, one may obtain information about 
the EOS parameters \citep{SNIO2012}.

\begin{figure}
\begin{center}
\begin{tabular}{cc}
\includegraphics[scale=0.5]{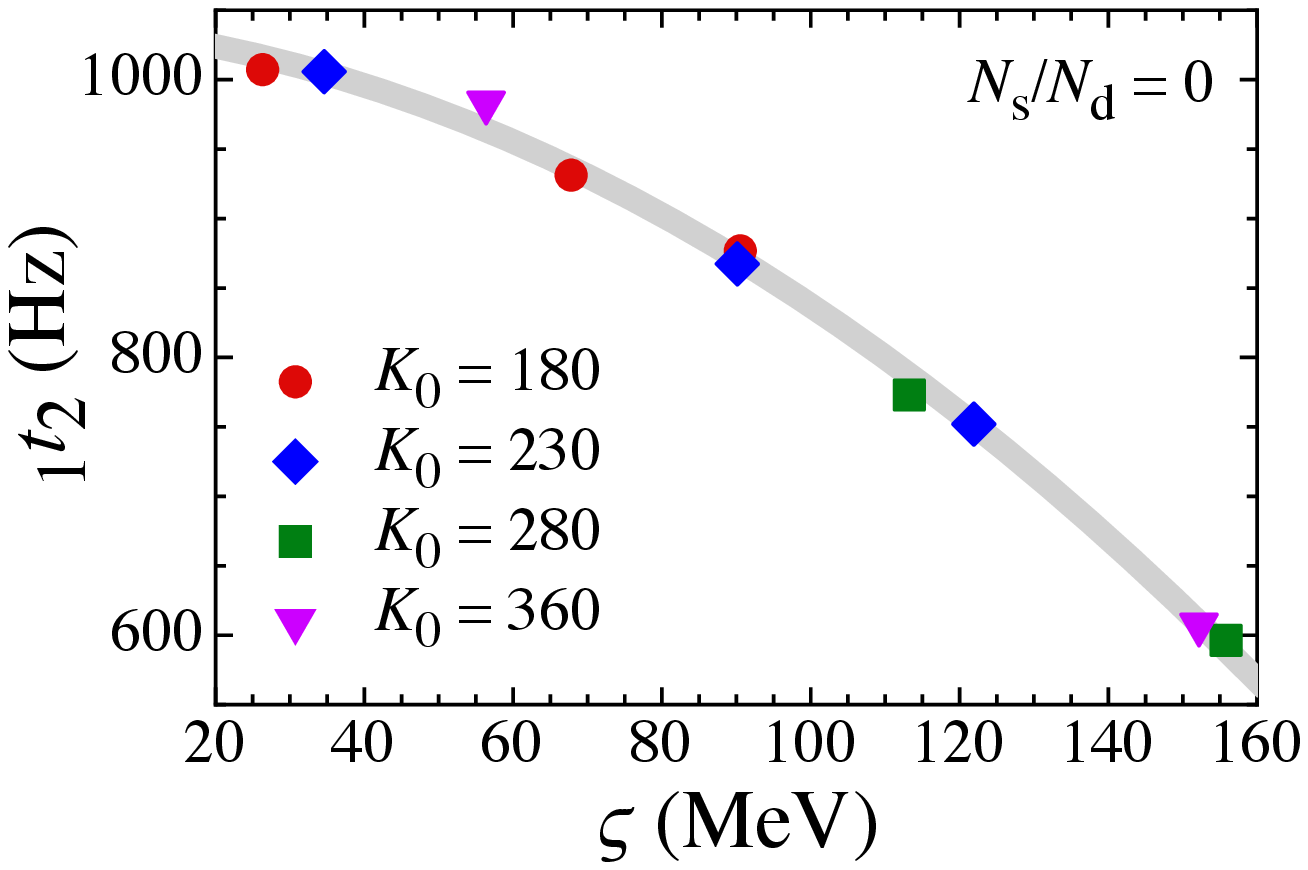} &
\includegraphics[scale=0.5]{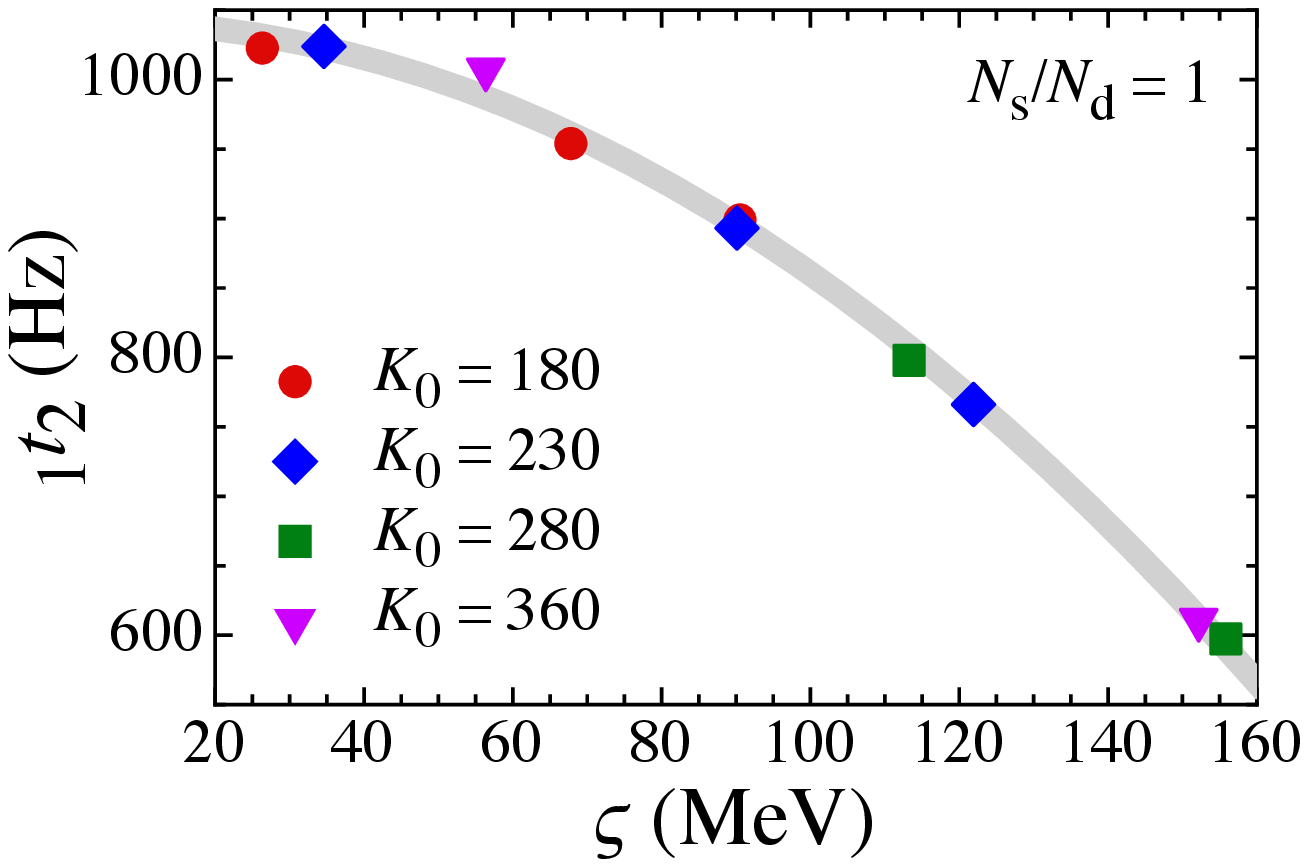} 
\end{tabular}
\end{center}
\caption{
The 1st overtone frequencies of the $\ell=2$ torsional oscillations, 
${}_1t_2$, calculated for various sets of the EOS parameters
and for a neutron star model with $M=1.4M_\odot$ and $R=12$ km
are plotted as a function of $\varsigma$ defined by Eq.\ (\ref{eq:ss}).  
The left and right panels correspond to the results with $N_s/N_d=0$ 
and 1, respectively, in the phase of cylindrical nuclei.  
In both panels, the thick solid line denotes the fitting formula 
given by Eq.\ (\ref{eq:1t2s}).
}
\label{fig:1t2-M14R12}
\end{figure}

It is of great use to find a parameter that is constructed by
a combination of $K_0$ and $L$ in such a way as to characterize the 
frequencies of the 1st overtones.  To this end, let us consider the 
combination $(K_0^iL^j)^{1/(i+j)}$ with integer numbers $i$ and $j$.   
As a result of the parameter search, we find out a suitable combination, 
i.e.,
\begin{equation}
  \varsigma = (K_0^4L^5)^{1/9}.  \label{eq:ss}
\end{equation}
We remark that the combination of $K_0$ and $L$ in $\varsigma$ is different 
from that in the parameter $\eta$ defined by $\eta=(K_0L^2)^{1/3}$, 
which is appropriate for describing the mass and radius of
low-mass neutron stars \citep{SIOO2014}.  In Fig.\ \ref{fig:1t2-M14R12}, 
the frequencies of the 1st overtones calculated for various EOS 
parameter sets are plotted as a function of $\varsigma$ by
adopting a typical neutron star model with $M=1.4M_\odot$ and 
$R=12$ km and setting $N_s/N_d=0$ and 1 in the phase of 
cylindrical nuclei.  From this figure, we find that the 
calculated frequencies behave smoothly as $\varsigma$ changes.

We can derive a fitting formula for the $\ell=2$ frequencies 
of the 1st overtones as the following quadratic function 
of $\varsigma$:
\begin{equation}
  {}_1t_2 = d_2^{(0)} + d_2^{(1)} \varsigma + d_2^{(2)} \varsigma^2,  \label{eq:1t2s}
\end{equation}
where $d_2^{(0)}$, $d_2^{(1)}$, and $d_2^{(2)}$ are adjustable coefficients
that depend on $M$ and $R$.  Formula (\ref{eq:1t2s}) well reproduces the
calculated frequencies as shown in Fig.\ \ref{fig:1t2-M14R12}. 
Additionally, we can confirm that the $\ell$-th frequencies of the 1st 
overtones calculated for various neutrons star models can also be well 
reproduced by a quadratic function of $\varsigma$,
\begin{equation}
  {}_1t_\ell = d_\ell^{(0)} + d_\ell^{(1)} \varsigma + d_\ell^{(2)} \varsigma^2,  \label{eq:1tls}
\end{equation}
where $d_\ell^{(0)}$, $d_\ell^{(1)}$, and $d_\ell^{(2)}$ are adjustable 
coefficients that depend on $M$ and $R$.

\begin{figure}
\begin{center}
\begin{tabular}{cc}
\includegraphics[scale=0.5]{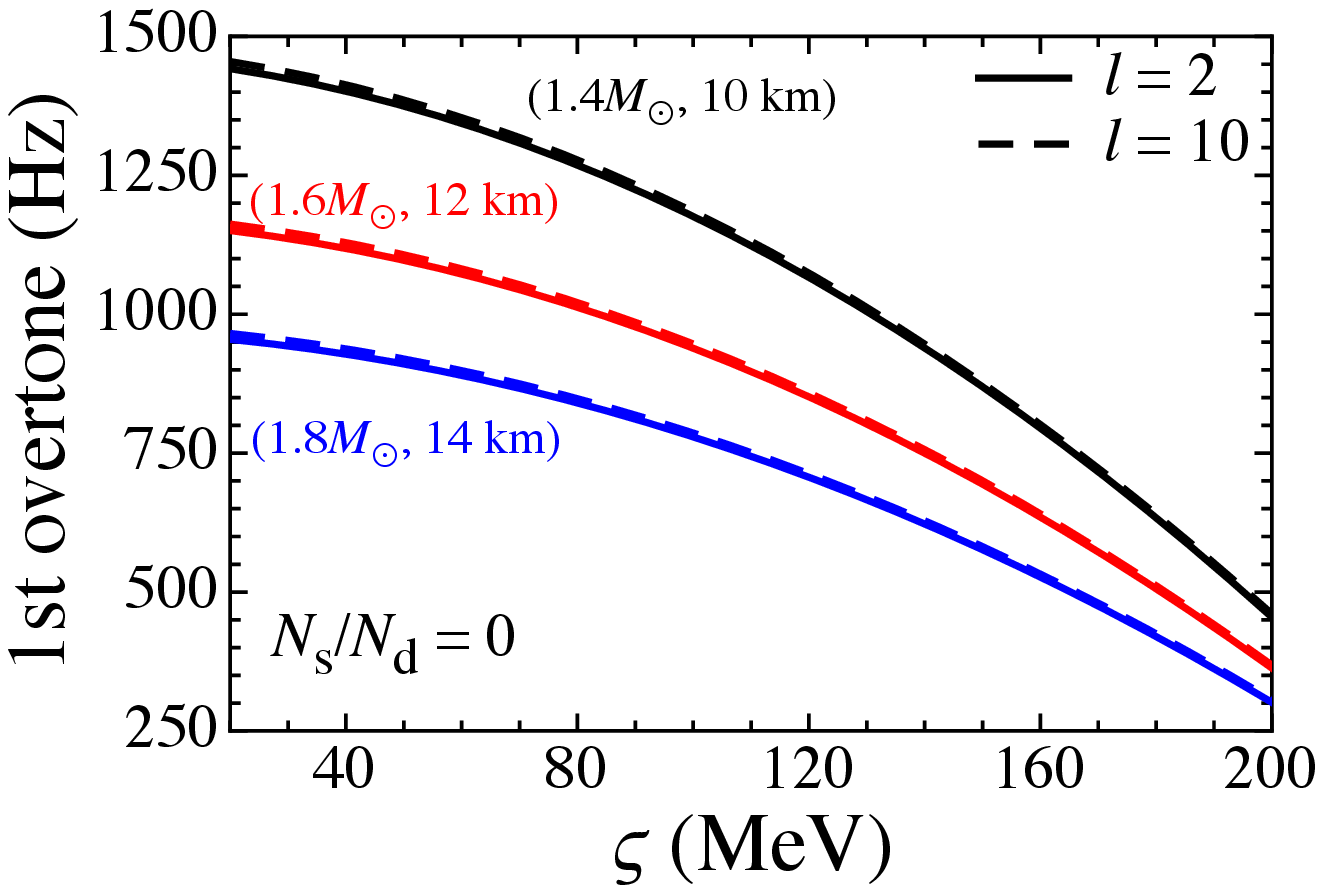} &
\includegraphics[scale=0.5]{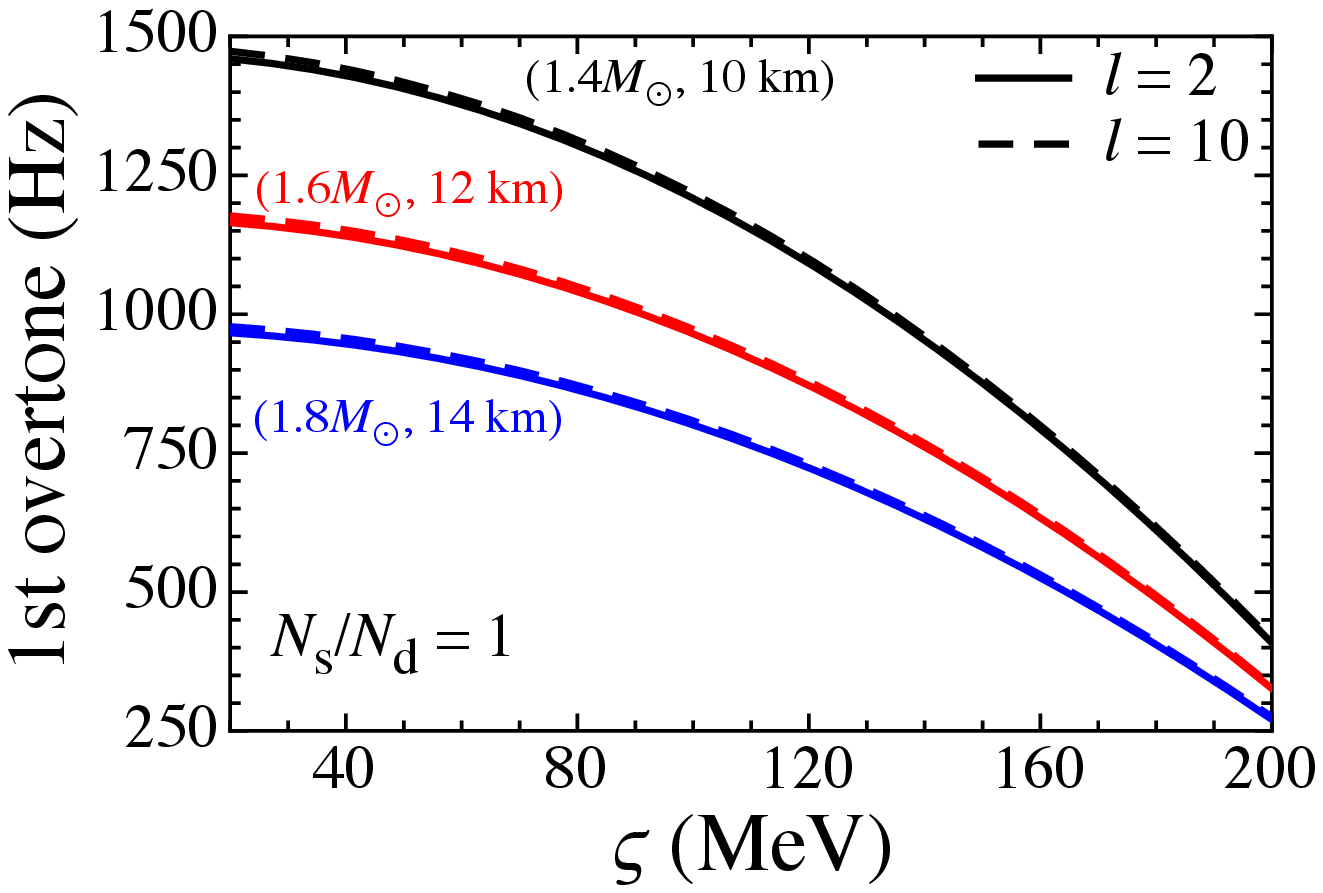} 
\end{tabular}
\end{center}
\caption{
The 1st overtones of the $\ell=2$ (solid lines) and 10 (dashed lines) 
torsional oscillations are shown as a function of $\varsigma$ in the
case of $N_s/N_d=0$ (left) and 1 (right) in the phase of 
cylindrical nuclei.  In each panel, the 
lines from top to bottom correspond to the fitting formula
[Eq. (\ref{eq:1tls})] adjusted to the results obtained for neutron 
star models with $(M, R)=(1.4M_\odot, 10\ {\rm km})$, 
$(1.6M_\odot, 12\ {\rm km})$, and $(1.8M_\odot, 14\ {\rm km})$.
}
\label{fig:1st-sigma}
\end{figure}

Unlike the fundamental frequencies of torsional oscillations, 
however, the overtone frequencies are closely 
spaced with respect to $\ell$ \citep{HC1980}, as is evident
from comparison of the $\ell=2$ and 10 1st overtone frequencies 
in Fig.\ \ref{fig:1st-sigma}.  Hereafter we thus focus 
only on the $\ell=2$ frequencies of the 1st overtones. 
Simultaneously, from this figure, one can observe that the 
frequencies of the 1st overtones strongly depend on the 
adopted neutron star models.  This dependence, which
occurs even with the compactness $M/R$ almost fixed, reflect 
the fact that the crust thickness $\Delta R$ scales as $R$
for fixed compactness $M/R$ \citep{SIO2017b}.

\begin{figure}
\begin{center}
\includegraphics[scale=0.5]{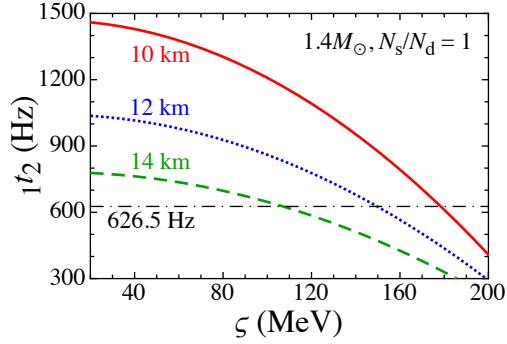} 
\end{center}
\caption{
The fitting formula [Eq. (\ref{eq:1t2s})] adjusted to the calculations of 
${}_1t_2$ obtained for $1.4M_\odot$ neutron stars of $R=10$ km (solid line), 
12 km (dotted line), and 14 km (dashed line) in the case of $N_s/N_d=1$ in 
the phase of cylindrical nuclei.  The 626.5 Hz QPO observed in SGR 1806$-$20 
is also shown as dot-dashed line.
}
\label{fig:1t2-M14}
\end{figure}

\begin{figure}
\begin{center}
\includegraphics[scale=0.5]{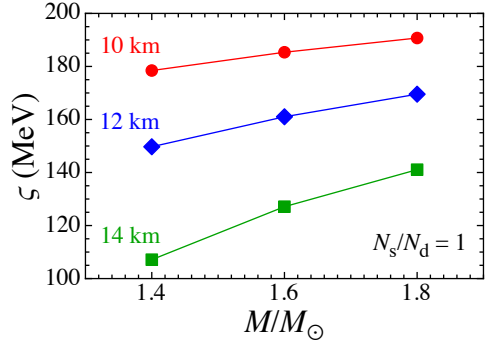} 
\end{center}
\caption{
 The optimal values of $\varsigma$ obtained in such a way as to 
explain the observed QPO 626.5 Hz in terms of ${}_1t_2$ are 
plotted for various neutron star models with $R=10, 12, 14$ km, 
$M=1.4M_\odot, 1.6M_\odot, 1.8 M_\odot$, and $N_s/N_d=1$ in the phase of 
cylindrical nuclei.
}
\label{fig:fit-s-Ns10}
\end{figure}

Now, we proceed to constrain $\varsigma$ by identifying the observed 
high frequency QPOs as the 1st overtones of crustal torsional 
oscillations.  Although most of the QPOs observed in SGR 1806--20 and in 
SGR 1900+14 have a central frequency lower than 160 Hz, the 626.5 and 
1837 Hz QPOs are also observed in SGR 1806--20.  Such high frequencies may 
come from something other than the torsional oscillations, e.g., 
polar type oscillations of neutron stars, but in the present study, we 
simply assume that the observed QPOs arise from purely crustal torsional 
oscillations.  Then, it is reasonable to identify the 626.5 Hz QPO as 
the 1st overtone of crustal torsional oscillations of some $\ell$.
By taking this identification for granted and by recalling that the calculated
1st overtone frequencies are closely spaced with respect to $\ell$, we 
can constrain the value of $\varsigma$ via comparison of the 
calculations of ${}_1t_2$ with 626.5 Hz as shown in Fig.\ \ref{fig:1t2-M14}.
From this figure, one can observe that the optimal values of 
$\varsigma$ are 178.5, 149.7, and 107.1 MeV for $1.4M_\odot$
neutron stars of $R=10$, 12, and 14 km, respectively, in the case of 
$N_s/N_d=1$ in the phase of cylindrical nuclei.  In a similar way, 
the optimal values of $\varsigma$ can be obtained for various
neutron star masses as shown in Fig.\ \ref{fig:fit-s-Ns10}.  We find 
that for each $R$, the optimal $\varsigma$ increases with $M$.  This is 
because $\Delta R$, which typically behaves as 
$\Delta R /R \simeq 2.1\times10^{-2} (R/M)(1-2M/R)$ \citep{SIO2017b}, decreases 
with the compactness $M/R$.

\begin{figure}
\begin{center}
\includegraphics[scale=0.5]{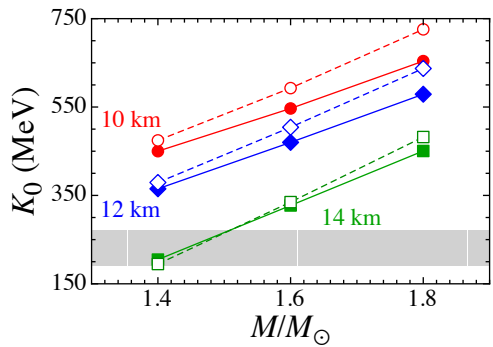} 
\end{center}
\caption{
The constraint on $K_0$ obtained by combining the constraints on $L$ and 
$\varsigma$ from the observed QPOs in SGR 1806--20 is plotted for 
various neutron star models with $N_s/N_d=1$ (filled marks with solid lines) 
and $N_s/N_d=0$ (open marks with dashed lines) in the phase of 
cylindrical nuclei. The painted region denotes the empirical 
constraint on $K_0$ \citep{KM2013}.
}
\label{fig:fit-K0-Ns10}
\end{figure}

Once we obtain constraints on $L$ and $K_0$, we can obtain information
about $M$ and $R$ of the compact object associated with the QPOs.  In fact,
for each neutron star model, we can derive the optimal value of 
$L$ in such a way as to explain the low-lying QPOs observed in 
SGR 1806--20 except the 26 Hz one as shown in Fig.\ 
\ref{fig:1806-Ns10} in the case of $N_s/N_d=1$ in the phase of 
cylindrical nuclei.  By combining this $L$ with the constraint on 
$\varsigma$ as shown in Fig.\ \ref{fig:fit-s-Ns10}, we can then 
constrain $K_0$ for each neutron star model via $K_0=(\varsigma^9/L^5)^{1/4}$.
The resultant values of $K_0$ are plotted in Fig.\ \ref{fig:fit-K0-Ns10}
with the filled marks and solid lines. 
On the other hand, the value of $K_0$ is constrained from experiments
on nuclear giant monopole resonances, e.g., $K_0=230\pm40$ MeV \citep{KM2013}.
We adopt this constraint on $K_0$ as a typical one although it is
still model dependent (see, e.g., \cite{SSM2014}).  This constraint on $K_0$,
which is also shown in Fig.\ \ref{fig:fit-K0-Ns10}, leads to
$M\simeq 1.4$--$1.5M_\odot$ for $R=14$ km, 
and presumably $M\simeq 1.2$--$1.4M_\odot$ for $R=13$ km via
extrapolations from the present calculations, 
as the favored neutron star models by the 
QPOs observed in SGR 1806--20 up to 626.5 Hz.  This suggests that a compact 
object in SGR 1806--20 would have a relatively low mass and large radius. 
A similar result can also be obtained in the case 
of $N_s/N_d=0$ in the phase of cylindrical nuclei, as shown in Fig.\ \ref{fig:fit-K0-Ns10}
with the open marks and dashed lines.
We remark in passing that a stellar model with still lower 
mass and smaller radius than that mentioned above might be acceptable
judging from Fig.\ \ \ref{fig:fit-K0-Ns10}, but 
such a model would lead to a larger value of the optimal $L$, which
in turn would be presumably inconsistent with the systematic analysis of
the mass-radius relation of low-mass neutron stars \citep{SIOO2014}.  
We also remark that the same kind of
constraint on the neutron star model for SGR 1806--20 was obtained from a 
rather restricted set of the EOS models for neutron star matter 
\citep{DSB2014}.

Given the present constraint on the neutron star model for 
SGR 1806--20, we can drastically improve the constraint on $L$ 
over that shown in Figs.\ \ref{fig:fit-Ns10} and \ref{fig:fit-Ns00}.
In fact, the $M$ and $R$ constraint mentioned above leads to the
favored values of $L$ by the QPOs observed in SGR 1806--20 up to 626.5 Hz
as roughly 62--73 (58--70) MeV for $N_s/N_d=1$ (0) in the phase
of cylindrical nuclei.  Within the uncertainty in this $N_s/N_d$, we
can finally obtain $L\simeq 58$--73 MeV, which seems consistent
with the existing empirical constraint on $L$ 
\citep{Tsang2012,L2014,Newton2014,BB2016}.  For these favored $L$ 
values, which are larger than 50 MeV, the pasta region is generally narrow.  
The cylindrical nuclei are thus expected to have only a little effect on the 
favored $L$ values.  In fact, the overtone frequencies ${}_1t_2$ calculated 
by ignoring the presence of cylindrical nuclei \citep{SNIO2012} are basically 
located between the present results obtained with $N_s/N_d=0$ and 1 in the 
phase of cylindrical nuclei.  We remark that the 26 Hz 
QPO can be explained in terms of the torsional oscillations 
in the phases of cylindrical-hole and spherical-hole nuclei, as 
mentioned in the previous subsection, while the 1837 Hz QPO may well be 
explained in terms of the overtones of the crustal torsional oscillations 
of some specific $n$ and $\ell$.
On the other hand, if a possible new finding of the QPO of
frequency less than 10 Hz pointed out by \cite{PGSE2018} is a true signal, 
the situation would become more complicated.  In fact, the new  
candidate QPO frequency is too low to be identified in 
terms of purely crustal torsional oscillations. Then, one might  
have to consider an alternative oscillation mechanism such 
as magnetic oscillations.

\section{Conclusion}
\label{sec:V}

We systematically calculate the fundamental and overtone eigenfrequencies 
of torsional oscillations that are excited in the crustal region 
composed of spherical and cylindrical nuclei by newly evaluating the 
shear modulus of the triangular lattice of cylindrical nuclei in such a way 
as to be consistent with the equilibrium configuration of neutron star matter
obtained from various sets of the OI-EOS.  
As a first step to interpret the observed QPO frequencies, we focus on shear oscillations inside the crust.  Despite the presence of the interaction of the shear oscillations with magnetic fields that penetrate the core, we prefer to neglect all related complications in this work and to improve on uncertainties associated with the calculations of purely crustal shear modes.  In fact,
the comprehensive study that includes
uncertainties in the EOS parameters $L$ and $K_0$, in the entrainment 
parameter $N_s/N_d$ in the phase of cylindrical nuclei, and in the
neutron star parameters $M$ and $R$ allows us to make a quantitative
comparison between the calculated eigenfrequencies and the observed QPO 
frequencies.  The resultant constraint on $L$ from the low frequency QPO data
is close to the previous constraint obtained by ignoring the effect of
nonzero shear modulus of the phase of cylindrical nuclei, but leads to 
a constraint on $M$ and $R$ via the high frequency QPO data and the empirical 
constraint on $K_0$.  This information on $M$ and $R$ helps us to further 
constrain $L$ as $\sim58$--73 MeV.

To make better estimates of the eigenfrequencies of crustal torsional
oscillations, we will have to additionally consider shell and pairing
effects on the charge number in the phase of spherical nuclei, the influence
of nonzero pairing gap on the entrainment effect, and the question of how
crustal polydispersity, magnetic fields, and plasticity affect the shear motion 
in a situation relevant for magnetars \citep{KP2015,vHL2011,L2016}.  
All these effects would definitely shift the pattern of the eigenfrequencies, 
leading to modification of the final constraint 
on $L$.  Moreover, all we can mention about the 26 Hz QPO observed in SGR 
1806--20 at present is the possibility that it arises from the fundamental
$\ell=2$ torsional mode in the region composed of cylindrical-hole and 
spherical-hole nuclei.  Whether or not this mode, if occurring in the 
deepest region of the crust, would be observable in the light curve has 
yet to be confirmed.

This work was supported in part by Grant-in-Aid for Scientific 
Research (C) through Grant No.\ 17K05458 provided by the
Japan Society for the Promotion of Science (JSPS) and in 
part by Grant-in-Aid for Scientific Research on Innovative 
Areas through No.\ 24105008 provided by the Ministry of 
Education, Culture, Sports, Science and Technology of Japan 
(MEXT).



\end{document}